\documentclass[prb,preprint]{revtex4-1} 

\usepackage{amsmath} 
\usepackage{amsfonts} 
\usepackage{graphicx} 
\usepackage{bm}
\usepackage{mathptmx}
\usepackage{amssymb} 
\usepackage{eucal}

\def\0{\hbox{\rm \bf 0}}
\def\1{\hbox{\rm \bf 1}}
\def\I{\hbox{\rm \bf I}}
\def\J{\hbox{\rm \bf J}}
\def\K{\hbox{\rm \bf K}}
\def\C{\hbox{\rm \bf i}}

\font\titlefont=cmssbx10 at 18pt
\font\authorfont=cmr12
\font\abstractfont=cmr10
\font\abstractit=cmti10

\font\sectionfont=cmssbx10 at 14pt

\font\reftitlefont=cmssbx10 at 12pt
\font\reftextfont=cmr10

\begin{document}

\vskip45pt
\begin{center}
{\titlefont
Biquaternion formulation of  \\
relativistic tensor dynamics\\
}

\vskip12pt
{\authorfont
E.P.J. de Haas (Paul) \\
Nijmegen, The Netherlands\\
Email: haas2u[AT]gmail.com}
\end{center}
\vskip18pt
{\abstractfont
In this paper we show how relativistic tensor dynamics and
relativistic electrodynamics can be formulated in a biquaternion
tensor language. The treatment is restricted to mathematical
physics, known facts as the Lorentz Force Law and the Lagrange
Equation are presented in a relatively new formalism. The goal is to
fuse anti-symmetric tensor dynamics, as used for example in
relativistic electrodynamics, and symmetric tensor dynamics, as used
for example in introductions to general relativity, into one single
formalism: a specific kind of biquaternion tensor calculus.}
\vskip12pt {\abstractfont {\abstractit Keywords}:
biquaternion, relativistic dynamics, Lorentz Force Law, Lagrange
Equation}

\vskip12pt
{\noindent \sectionfont Introduction}
\vskip6pt

We start by quoting Yefremov. {\it One can say that space-time model
and kinematics of the Quaternionic Relativity are nowadays studied
in enough details and can be used as an effective mathematical tool
for calculation of many relativistic effects. But respective
relativistic dynamic has not been yet formulated, there are no
quaternionic field theory; Q-gravitation, electromagnetism, weak and
strong interactions are still remote projects. However, there is a
hope that it is only beginning of a long way, and the theory will
mature. [1]}

We hope that the content of this paper will contribute to the
project described by Yefremov.

Quaternions can be represented by the basis $(\1,\I,\J,\K)$. This
basis has the properties $\I\I=\J\J=\K\K=-\1$; $\1\1=\1$;
$\1\K=\K\1=\K$ for $\I, \J, \K$; $\I\J=-\J\I=\K$; $\J\K=-\K\J=\I$;
$\K\I=-\I\K=\J$. A quaternion number in its summation representation
is given by $A = a_0\1 + a_1\I + a_2\J + a_3\K $, in which the
$a_{\mu}$ are real numbers . Biquaternions or complex quaternions in
their summation representation are given by
\begin{eqnarray}
\nonumber C = A + \C B = \\
\nonumber (a_0+ \C b_0)\1 + (a_1+ \C b_1)\I + (a_2+ \C
b_2)\J + (a_3+ \C b_3)\K  = \\
a_0\1 + a_1\I + a_2\J + a_3\K + \C b_0\1 + \C b_1\I + \C b_2\J + \C
b_3\K,
\end{eqnarray}
in which the $c_{\mu}=a_{\mu}+ \C b_{\mu} $ are complex numbers and
the $a_{\mu}$ and $b_{\mu}$ are real numbers. The complex conjugate
of a biquaternion $C$ is given by $\widetilde{C} = A - \C B$. The
quaternion conjugate of a biquaternion is given by
\begin{eqnarray}
\nonumber C^{\dag} = A^{\dag} + \C B^{\dag} = \\
(a_0+ \C b_0)\1 - (a_1+ \C b_1)\I - (a_2+ \C b_2)\J - (a_3+ \C
b_3)\K.
\end{eqnarray}
In this paper we only use the complex conjugate of biquaternions.

Biquaternions or complex quaternions in their vector representation
are given by
\begin{equation}
C_{\mu}= \left[%
\begin{array}{c}
  c_0 \1\\
  c_1 \I\\
  c_2 \J\\
  c_3 \K\\
\end{array}%
\right],
\end{equation}
or by
\begin{equation}
C^{\mu}= \left[%
  c_0 \1,
  c_1 \I,
  c_2 \J,
  c_3 \K
\right]
\end{equation}

We apply this to the space-time four vector of relativistic
biquaternion 4-space $R_{\mu}$ as
\begin{equation}
R_{\mu}= \left[%
\begin{array}{c}
  \C ct \1\\
  r_1 \I\\
  r_2 \J\\
  r_3 \K\\
\end{array}%
\right]=
\left[%
\begin{array}{c}
  \C r_0 \1\\
  r_1 \I\\
  r_2 \J\\
  r_3 \K\\
\end{array}%
\right].
\end{equation}
The space-time distance $s$ can be defined as
$\widetilde{R}^{\mu}R^{\mu}$, or
\begin{equation}
\widetilde{R}^{\mu}R^{\mu}=\left[%
  - \C ct \1,
  r_1 \I,
  r_2 \J,
  r_3 \K
\right]\left[%
  \C ct \1,
  r_1 \I,
  r_2 \J,
  r_3 \K
\right],
\end{equation}
giving
\begin{equation}
\widetilde{R}^{\mu}R^{\mu}=
  c^2 t^2 \1 -
  r_1^2 \1 -
  r_2^2 \1 -
  r_3^2 \1 = ( c^2 t^2 - r_1^2 - r_2^2 - r_3^2)\1 .
\end{equation}
So we get $\widetilde{R}^{\mu}R^{\mu}= s \1$ with the usual
\begin{equation}
s = c^2 t^2 - r_1^2 - r_2^2 - r_3^2=r_0^2- r_1^2 - r_2^2 - r_3^2
\end{equation}
providing us with a $(+,-,-,-)$ signature.

\vskip12pt {\noindent
\sectionfont Adding the dynamic vectors}
\vskip6pt

The basic definitions we use are quite common in the usual
formulations of relativistic dynamics, see [2], [3]. We start with
an observer who has a given three vector velocity as ${\bf v}$, a
rest mass as $m_0$ and an inertial mass $m_i=\gamma m_0$, with the
usual $\gamma = (\sqrt{1-v^2/c^2})^{-1}$. We use the Latin suffixes
as abbreviations for words, not for numbers. So $m_i$ stands for
inertial mass and $U_p$ for potential energy. The Greek suffixes are
used as indicating a summation over the numbers 0, 1, 2 and 3. So
$P_{\mu}$ stands for a momentum four-vector with components
$p_0=\frac{1}{c}U_i$, $p_1$, $p_2$ and $p_3$. The momentum
three-vector is written as ${\bf p}$ and has components $p_1$, $p_2$
and $p_3$.

We define the coordinate velocity four vector as
\begin{equation}
 V_{\mu}= \frac{d}{d t}R_{\mu}=
  \left[%
\begin{array}{c}
  \C c \1\\
  v_1 \I\\
  v_2 \J\\
  v_3 \K\\
\end{array}%
\right] =
  \left[%
\begin{array}{c}
  \C v_0 \1\\
  v_1 \I\\
  v_2 \J\\
  v_3 \K\\
\end{array}%
\right].
\end{equation}
The proper velocity four vector on the other hand will be defined
using the proper time $t_0$, with $t=\gamma t_0$, as
\begin{equation}
 U_{\mu}= \frac{d}{d t_0}R_{\mu}=
 \frac{d}{\frac{1}{\gamma}d t}R_{\mu}=
 \gamma V_{\mu}=
  \left[%
\begin{array}{c}
   \C \gamma c \1\\
  \gamma v_1 \I\\
  \gamma v_2 \J\\
  \gamma v_3 \K\\
\end{array}%
\right].
\end{equation}
The momentum four vector will be
\begin{equation}
 P_{\mu}= m_i V_{\mu}= m_0 U_{\mu}.
\end{equation}

We further define the rest mass density as
\begin{equation}
    \rho_0 = \frac{d m_0}{d V_0},
\end{equation}
so with
\begin{equation}
    d V = \frac{1}{\gamma}d V_0
\end{equation}
and the inertial mass density as
\begin{equation}
    \rho_i = \frac{d m_i}{d V}
\end{equation}
we get, in accordance with Arthur Haas' 1930 exposition on
relativity ([4], p. 365),
\begin{equation}
    \rho_i = \frac{d m_i}{d V}= \frac{d \gamma
    m_0}{\frac{1}{\gamma}d V_0}= \gamma^2 \rho_0.
\end{equation}

The momentum density four vector will be defined as
\begin{equation}
 G_{\mu}=
  \left[%
\begin{array}{c}
  \C\frac{1}{c}u_i \1 \\
  g_1 \I\\
  g_2 \J\\
  g_3 \K\\
\end{array}%
\right] =
  \left[%
\begin{array}{c}
  \C g_0 \1 \\
  g_1 \I\\
  g_2 \J\\
  g_3 \K\\
\end{array}%
\right],
\end{equation}
in which we used the inertial energy density $u_i=\rho_i c^2$.
For this momentum density four vector we have the variations
\begin{equation}
G_{\mu}=\frac{d}{d V}P_{\mu} = \frac{d m_i}{d V}V_{\mu} = \rho_i
V_{\mu} = \gamma^2 \rho_0 V_{\mu}= \gamma \rho_0 U_{\mu} = \gamma
G_{\mu}^{proper}.
\end{equation}

The four vector partial derivative $\partial_{\mu}$ will be defined
as
\begin{equation}
 \partial_{\mu}=
  \left[%
\begin{array}{c}
  - \C\frac{1}{c}\partial_t \1 \\
  \nabla_1 \I\\
  \nabla_2 \J\\
  \nabla_3 \K\\
\end{array}%
\right]\equiv \frac{\partial}{\partial R_{\mu}}.
\end{equation}

The electrodynamic potential four vector will be defined as
\begin{equation}
 A_{\mu}=
  \left[%
\begin{array}{c}
  \C\frac{1}{c}\phi \1 \\
  A_1 \I\\
  A_2 \J\\
  A_3 \K\\
\end{array}%
\right] =
  \left[%
\begin{array}{c}
  \C A_0 \1 \\
  A_1 \I\\
  A_2 \J\\
  A_3 \K\\
\end{array}%
\right].
\end{equation}

The electric four current density will be given by
\begin{equation}
 J_{\mu}=
  \left[%
\begin{array}{c}
  \C c \rho_e \1 \\
  J_1 \I\\
  J_2 \J\\
  J_3 \K\\
\end{array}%
\right] =
  \left[%
\begin{array}{c}
  \C J_0 \1 \\
  J_1 \I\\
  J_2 \J\\
  J_3 \K\\
\end{array}%
\right]= \rho_e V_{\mu},
\end{equation}
with $\rho_e$ as the electric charge density.

\vskip12pt {\noindent
\sectionfont Adding the dynamic vector products, scalars}
\vskip6pt

The dynamic Lagrangian density $\mathcal{L}$ can be defined as
\begin{equation}
\mathcal{L}=-\widetilde{V}^{\nu}G^{\nu}= -(u_i - {\bf v}\cdot {\bf g})\1 = -
u_0 \1
\end{equation}
and the accompanying Lagrangian $L$ as
\begin{equation}
L=-\widetilde{V}^{\nu}P^{\nu}= -(U_i - {\bf v}\cdot {\bf p})\1 = -
\frac{1}{\gamma}U_0 \1,
\end{equation}
with $u_0$ as the rest system inertial energy density and $U_0$ as
the rest system inertial energy. The latter is the usual Lagrangian
of a particle moving freely in empty space.

The Lagrangian density of a massless electric charge density current
in an electrodynamic potential field can be defined as
\begin{equation}
\mathcal{L}=-\widetilde{J}^{\nu}A^{\nu}= -(\rho_e \phi - {\bf J}\cdot {\bf
A})\1.
\end{equation}

On the basis of the Lagrangian density we can define a four force
density as
\begin{equation}\label{RFD1}
f_{\mu}\equiv \frac{\partial \mathcal{L}}{\partial R_{\mu}}= \partial_{\mu}
\mathcal{L} = -\partial_{\mu}u_0.
\end{equation}
In the special case of a static electric force field, and without
the densities, the field energy is $U_0=q\phi_0$ and the
relativistic force reduces to the Coulomb Force
\begin{equation}
{\bf F}= - \nabla U_0 = -q \nabla \phi_0.
\end{equation}
Using $\mathcal{L} = -\widetilde{V}^{\nu}G^{\nu}$ the relativistic four force
density of Eq.(\ref{RFD1}) can be written as
\begin{equation}
f_{\mu}= -\partial_{\mu}\widetilde{V}^{\nu}G^{\nu}.
\end{equation}

We can define the absolute time derivative $\frac{d}{d t}$ of a
continuous, perfect fluid like, space/field quantity through
\begin{equation}
  -V^{\mu}\widetilde{\partial}^{\mu}=-\widetilde{V}^{\mu}\partial^{\mu}= {\bf v}\cdot \nabla + \partial_t\1
 =  \frac{d}{d t}\1.
\end{equation}
Thus we can define the mechanic four force density as
\begin{equation}
f_{\mu}\equiv \frac{d}{d t}G_{\mu}=
-(V^{\nu}\widetilde{\partial}^{\nu})G_{\mu}=
-V^{\nu}(\widetilde{\partial}^{\nu}G_{\mu}),
\end{equation}
using the fact that biquaternion multiplication is associative.

\vskip12pt {\noindent
\sectionfont Adding the dynamic vector products, tensors}
\vskip6pt

The mechanical stress energy tensor, introduced by Max von Laue in
1911, was defined by him as ([5], [6], p.150)
\begin{equation}\label{UU}
T^{\nu}_{\hspace{0.08in} \mu}= \rho_0 U^{\nu}U_{\mu}.
\end{equation}
Pauli gave the same definition in his standard work on relativity
([2], p. 117). With the vector and density definitions that we have
given we get
\begin{equation}
T^{\nu}_{\hspace{0.08in} \mu}= \rho_0 U^{\nu}U_{\mu}= \gamma^2
\rho_0 V^{\nu}V_{\mu}= \rho_i V^{\nu}V_{\mu}=V^{\nu}\rho_i
V_{\mu}=V^{\nu}G_{\mu}.
\end{equation}
So the mechanical stress energy tensor can also be written as
\begin{equation}\label{VG}
T^{\nu}_{\hspace{0.08in} \mu}=V^{\nu}G_{\mu}.
\end{equation}
In the exposition on relativity of Arthur Haas, the first definition
$ \rho_0 U^{\nu}U_{\mu}$ is described as the "Materie-tensor" of
General Relativity, while $\rho_i V^{\nu}V_{\mu}$ is described as
the "Materie-tensor" of Special Relativity ([4], p. 395 and p. 365).

Although the derivation seems to demonstrate an equivalence between
the two formulations of equation (\ref{UU}) and equation (\ref{VG}),
the difference between the two is fundamental. Equation (\ref{UU})
is symmetric by definition, while equation (\ref{VG}) can be
asymmetric, because, as von Laue already remarked in 1911, $V^{\nu}$
and $G_{\mu}$ do not have to be parallel all the time ([5], [6] p.
167) This crucial difference between $\rho_0 U^{\nu}U_{\mu}$ and
$V^{\nu}G_{\mu}$ was also discussed by de Broglie in connection with
his analysis of electron spin ([7], p. 55). In our context, where we
want to fuse the symmetric and antisymmetric formalism into one, we
prefer the stress energy density tensor of equation (\ref{VG}), the
one called the "Materie-tensor" of Special Relativity by Arthur
Haas.

So the stress energy density tensor $T^{\nu}_{\hspace{0.08in} \mu}$
can be given as
$T^{\nu}_{\hspace{0.08in}\mu}=\widetilde{V}^{\nu}G_{\mu}$ and gives
\begin{eqnarray}
\nonumber T^{\nu}_{\hspace{0.08in}\mu}=\left[- \C v_0 \1, v_1 \I,
v_2 \J, v_3 \K \right]
\left[%
\begin{array}{c}
  \C g_0 \1 \\
  g_1 \I\\
  g_2 \J\\
  g_3 \K\\
\end{array}%
\right]=\\
\left[%
  \begin{array}{cccc}
   v_0 g_0 \1         & \C  v_{1}g_0\I   & \C  v_{2}g_0\J & \C  v_{3}g_0\K  \\
  -\C v_0 g_{1}\I   & -v_{1}g_{1}\1             & -v_{2}g_{1}\K           &  v_{3}g_{1}\J   \\
  -\C v_0 g_{2}\J   &  v_{1}g_{2}\K             & -v_{2}g_{2}\1           & -v_{3}g_{2}\I   \\
  -\C v_0 g_{3}\K   & -v_{1}g_{3}\J             &  v_{2}g_{3}\I           & -v_{3}g_{3}\1   \\
  \end{array}
  \right]
\end{eqnarray}
Its trace is $T^{\nu\nu}=\widetilde{V}^{\nu}G^{\nu}= -\mathcal{L}$.

In relativistic dynamics we have a usual force density definition
through the four derivative of the stress energy density tensor
\begin{equation}
\partial^{\nu} T^{\nu}_{\hspace{0.08in}\mu}= -f_{\mu}
\end{equation}
or
\begin{equation}
\partial^{\nu} V^{\nu}G_{\mu}= -f_{\mu}
\end{equation}
We want to find out if these equations still hold in our
biquaternion version of the four vectors, tensors and their
products.

We calculate the left hand side and get for
$\partial^{\nu}T^{\nu}_{\hspace{0.08in}\mu}=\partial^{\nu}
\widetilde{V}^{\nu}G_{\mu}$:

\begin{eqnarray}
\nonumber \left[-\frac{\C}{c}\partial_t \1, \nabla_1 \I, \nabla_2
\J, \nabla_3 \K \right]\\
\left[%
  \begin{array}{cccc}
   v_0g_0 \1         &  \C v_1g_0\I              & \C v_2g_0\J              &\C v_3g_0\K  \\
  -\C v_0 g_1\I   & -v_1g_1\1             & -v_2g_1\K           &  v_3g_1\J   \\
  -\C v_0 g_2\J   &  v_1g_2\K             & -v_2g_2\1           & -v_3g_2\I   \\
  -\C v_0 g_3\K   & -v_1g_3\J             &  v_2g_3\I           & -v_3g_3\1   \\
  \end{array}
  \right]
\end{eqnarray}

which equals

\begin{eqnarray}
\nonumber \left[%
  \begin{array}{c}
   -\C\frac{1}{c}\partial_t v_0g_0 \1   - \C \nabla_1 v_{1}g_0\1          - \C\nabla_2  v_{2}g_0\1           - \C\nabla_3v_{3}g_0\1  \\
  -\frac{1}{c}\partial_t v_0 g_1\I   -\nabla_1 v_{1}g_{1}\I              -\nabla_2 v_{2}g_{1}\I           - \nabla_3 v_{3}g_{1}\I   \\
  -\frac{1}{c}\partial_t v_0 g_2\J   - \nabla_1 v_{1}g_{2}\J             -\nabla_2 v_{2}g_{2}\J            -\nabla_3 v_{3}g_{2}\J   \\
  -\frac{1}{c}\partial_t v_0 g_3\K    -\nabla_1 v_{1}g_{3}\K             - \nabla_2 v_{2}g_{3}\K            -\nabla_3 v_{3}g_{3}\K   \\
  \end{array}
  \right]\\
= - \left[%
  \begin{array}{c}
   \C(\frac{1}{c}\partial_t v_0 g_0 + \nabla_1 v_{1}g_0 + \nabla_2 v_{2}g_0 + \nabla_3 v_{3}g_0)\1  \\
  (\frac{1}{c}\partial_t v_0 g_{1} +\nabla_1 v_{1}g_{1}  +\nabla_2 v_{2}g_{1}   + \nabla_3 v_{3}g_{1})\I   \\
  (\frac{1}{c}\partial_t v_0 g_{2} +\nabla_1 v_{1}g_{2}  +\nabla_2 v_{2}g_{2}   +\nabla_3 v_{3}g_{2})\J   \\
  (\frac{1}{c}\partial_t v_0 g_{3} +\nabla_1 v_{1}g_{3}  + \nabla_2 v_{2}g_{3}  +\nabla_3 v_{3}g_{3})\K   \\
  \end{array}
  \right]
\end{eqnarray}

Using the chain rule this leads to
\begin{eqnarray}
  \nonumber \partial^{\nu}T^{\nu}_{\hspace{0.08in}\mu}=
- \left[%
  \begin{array}{c}
   \C(\frac{1}{c}v_0\partial_t  g_0 + v_{1}\nabla_1 g_0 + v_{2}\nabla_2 g_0 + v_{3}\nabla_3 g_0)\1  \\
  (\frac{1}{c}v_0\partial_t  g_{1} +v_{1}\nabla_1 g_{1}  +v_{2}\nabla_2 g_{1}   + v_{3}\nabla_3 g_{1})\I   \\
  (\frac{1}{c}v_0\partial_t  g_{2} +v_{1}\nabla_1 g_{2}  +v_{2}\nabla_2 g_{2}   +v_{3}\nabla_3 g_{2})\J   \\
  (\frac{1}{c}v_0\partial_t  g_{3} +v_{1}\nabla_1 g_{3}  + v_{2}\nabla_2 g_{3}  +v_{3}\nabla_3 g_{3})\K   \\
  \end{array}
  \right] - \\
    \nonumber
 \left[%
  \begin{array}{c}
   \C(\frac{1}{c}(\partial_t v_0) g_0 +(\nabla_1 v_{1})g_0    +(\nabla_2 v_2)g_0    +(\nabla_3 v_{3})g_0)\1  \\
  (\frac{1}{c}(\partial_t v_0 g_{1})  +(\nabla_1 v_{1})g_{1}  +(\nabla_2 v_2)g_{1}  +(\nabla_3 v_{3})g_{1})\I   \\
  (\frac{1}{c}(\partial_t v_0 g_{2})  +(\nabla_1 v_{1})g_{2}  +(\nabla_2 v_2)g_{2}  +(\nabla_3 v_{3})g_{2})\J   \\
  (\frac{1}{c}(\partial_t v_0 g_{3})  +(\nabla_1 v_{1})g_{3}  +(\nabla_2 v_2)g_{3}  +(\nabla_3 v_{3})g_{3})\K   \\
  \end{array}
  \right]\\
  \nonumber =
- (\frac{1}{c}v_0\partial_t  + v_1\nabla_1  + v_2\nabla_2 +
v_{3}\nabla_3)
\left[%
  \begin{array}{c}
   \C g_0\1  \\
  g_1\I   \\
  g_2\J   \\
  g_3\K   \\
  \end{array}
  \right]  \\
    \nonumber -(\frac{1}{c}\partial_t v_0 +\nabla_1 v_{1} +\nabla_2 v_2  +\nabla_3 v_{3})
 \left[%
  \begin{array}{c}
   \C g_0\1  \\
  g_{1}\I   \\
  g_{2}\J   \\
  g_{3}\K   \\
  \end{array}
  \right]\\
   = (\widetilde{V}^{\nu}\partial^{\nu})G_{\mu}+(\partial^{\nu}\widetilde{V}^{\nu})G_{\mu}
\end{eqnarray}
This can be abbreviated to
\begin{equation}
    \partial^{\nu} (\widetilde{V}^{\nu}G_{\mu})= (\widetilde{V}^{\nu}\partial^{\nu})
    G_{\mu}+ (\partial^{\nu}\widetilde{V}^{\nu})G_{\mu}
\end{equation}
So
\begin{equation}
   \partial^{\nu}T^{\nu}_{\hspace{0.08in}\mu}= (\widetilde{V}^{\nu}\partial^{\nu})
    G_{\mu}+ (\partial^{\nu}\widetilde{V}^{\nu})G_{\mu}.
\end{equation}
We have $\widetilde{V}^{\nu}\partial^{\nu}=-\frac{d}{dt}$ and if we
assume the bare particle velocity continuity equation
$\partial^{\nu}\widetilde{V}^{\nu}=0$, then we get
\begin{equation}
    \partial^{\nu}T^{\nu}_{\hspace{0.08in}\mu} = -\frac{d}{d
    t}G_{\mu}=-f_{\mu}.
\end{equation}

\vskip12pt {\noindent
\sectionfont Electrodynamic vector products}
\vskip6pt

If we apply this to the case in which we have a purely
electromagnetic four momentum density $G_{\mu}= \rho_e A_{\mu}$ then
we have
\begin{equation}
\mathcal{L} = - \widetilde{V}^{\nu}G^{\nu}= - \widetilde{V}^{\nu}\rho_e
A^{\nu}= - \widetilde{J}^{\nu}A^{\nu},
\end{equation}
and
\begin{equation}
T^{\nu}_{\hspace{0.08in}\mu}=\widetilde{V}^{\nu}G_{\mu}=\widetilde{J}^{\nu}A_{\mu}.
\end{equation}
The relativistic force equation
\begin{equation}
   \partial^{\nu}T^{\nu}_{\hspace{0.08in}\mu}= (\widetilde{V}^{\nu}\partial^{\nu})
    G_{\mu}+ (\partial^{\nu}\widetilde{V}^{\nu})G_{\mu}.
\end{equation}
can be given its electrodynamic expression as
\begin{equation}
   \partial^{\nu}T^{\nu}_{\hspace{0.08in}\mu}= (\widetilde{J}^{\nu}\partial^{\nu})
    A_{\mu}+ (\partial^{\nu}\widetilde{J}^{\nu})A_{\mu}.
\end{equation}
If the charge density current continuity equation
$\partial^{\nu}\widetilde{J}^{\nu}=0$ can be applied, then this
reduces to
\begin{equation}\label{DTJA}
   \partial^{\nu}T^{\nu}_{\hspace{0.08in}\mu}= (\widetilde{J}^{\nu}\partial^{\nu})
    A_{\mu}= (J^{\nu}\widetilde{\partial}^{\nu})A_{\mu}=J^{\nu}(\widetilde{\partial}^{\nu}A_{\mu}) .
\end{equation}

The electrodynamic force field tensor $B^{\nu}_{\hspace{0.08in}\nu}$
is given by
\begin{equation}\label{BDA}
B^{\nu}_{\hspace{0.08in}\mu}=\widetilde{\partial}^{\nu}A_{\mu}.
\end{equation}
In detail this reads
\begin{eqnarray}
\nonumber B^{\nu}_{\hspace{0.08in}\mu}=\left[\C\frac{1}{c}\partial_t
\1, \nabla_1 \I, \nabla_2 \J, \nabla_3 \K \right]
\left[%
\begin{array}{c}
  \C \frac{1}{c}\phi \1 \\
  A_1 \I\\
  A_2 \J\\
  A_3 \K\\
\end{array}%
\right]=\\
\left[%
  \begin{array}{cccc}
  - \frac{1}{c^2}\partial_t \phi\1         & \C  \frac{1}{c}\nabla_1\phi\I   & \C  \frac{1}{c}\nabla_2\phi\J & \C \frac{1}{c} \nabla_3\phi\K  \\
  \C \frac{1}{c}\partial_t A_1\I   & -\nabla_1 A_1\1             & -\nabla_2A_1\K           &  \nabla_3A_1\J   \\
  \C \frac{1}{c}\partial_t A_2\J   &  \nabla_1 A_2\K             & -\nabla_2A_2\1           & -\nabla_3A_2\I   \\
  \C \frac{1}{c}\partial_t A_3\K   & -\nabla_1 A_3\J             &  \nabla_2A_3\I           & -\nabla_3A_3\1   \\
  \end{array}
  \right]
\end{eqnarray}
To see that this tensor leads to the usual EM force field
biquaternion, we have to rearrange the tensor terms according to
their biquaternionic affiliation, so arrange them according to the
basis $(\1,\I,\J,\K)$. This results in
\begin{equation}
\left[%
    \begin{array}{c}
(- \frac{1}{c^2}\partial_t \phi-\nabla_1 A_1-\nabla_2A_2-\nabla_3A_3)\1   \\
(\nabla_2A_3-\nabla_3A_2 +\C \frac{1}{c} \nabla_1\phi + \C
\frac{1}{c}\partial_t A_1)\I \\
(\nabla_3 A_1-\nabla_1A_3 + \C \frac{1}{c} \nabla_2\phi + \C
\frac{1}{c}\partial_t A_2)\J \\
(\nabla_1A_2-\nabla_2A_1 + \C\frac{1}{c}\nabla_3\phi + \C\frac{1}{c}\partial_tA_3)\K\\
    \end{array}
  \right]
\end{equation}
This equals
\begin{equation}
F_{\mu}=
\left[%
    \begin{array}{c}
\widetilde{\partial}^{\nu}A^{\nu}\1   \\
(B_1 - \C \frac{1}{c}E_1)\I \\
(B_2 - \C \frac{1}{c}E_1)\J \\
(B_3 - \C \frac{1}{c}E_1)\K\\
    \end{array}
  \right]=
\left[%
    \begin{array}{c}
F_0 \1   \\
F_1\I \\
F_2 \J \\
F_3 \K\\
    \end{array}
  \right].
\end{equation}
For this biquaternion to be the exact match with the standard EM
force field, one has to add the Lorenz gauge condition
$F_0=\widetilde{\partial}^{\nu}A^{\nu}=0$. (If $F_0\neq 0$, then the
usual biquaternion expressions for the Lorentz Force and the two
inhomogenious Maxwell Equations contain extra terms. The
biquaternion formalism demonstrated in this paper doesn't involve
these extra terms.) The operation of rearranging the tensor terms
according to their biquaternion affiliation is external to the
mathematical physics of this paper. We try to develop a biquaternion
version of relativistic tensordynamics. The above operation destroys
the tensor arrangement of the terms involved. It is alien to the
system we try to develop in this context. It may be a useful
operation in others areas though, for example in quantum physics.

The electrodynamic force field tensor $B_{\mu}^{\hspace{0.08in}\nu}$
can also be given by
\begin{equation}
B_{\mu}^{\hspace{0.08in}\nu}=\widetilde{\partial}_{\mu}A^{\nu}.
\end{equation}
This leads to the same EM force field biquaternion.

The combination of Eq.(\ref{DTJA}) and Eq.(\ref{BDA}) leads to
\begin{equation}\label{TB}
   \partial^{\nu}T^{\nu}_{\hspace{0.08in}\mu}=
   J^{\nu}B^{\nu}_{\hspace{0.08in}\mu},
\end{equation}
which is valid if charge is conserved so if
$\partial^{\nu}\widetilde{J}^{\nu}=0$.

We can write Eq.(\ref{DTJA}) also as
\begin{equation}\label{TqE}
   \partial^{\nu}T^{\nu}_{\hspace{0.08in}\mu}= (\widetilde{J}^{\nu}\partial^{\nu})
    A_{\mu}= \rho_e(\widetilde{V}^{\nu}\partial^{\nu})A_{\mu}= - \rho \frac{d}{dt}A_{\mu}.
\end{equation}

The two EM force expression we gave in this and the previous
sections based on
$f_{\mu}=-\partial^{\nu}T^{\nu}_{\hspace{0.08in}\mu}$ and
$f_{\mu}=\partial_{\mu}\mathcal{L}$ do not result in the well known Lorentz
Force. But we can establish a relationship between these force
expressions and the Lorentz Force.

\vskip12pt {\noindent
\sectionfont The Lorentz Force Law}
\vskip6pt

The relativistic Lorentz Force Law in its density form is given by
the expression
\begin{equation}\label{LFL}
f_{\mu}=   J^{\nu}(\widetilde{\partial}^{\nu}A_{\mu}) -
(\partial_{\mu}\widetilde{A}^{\nu})J^{\nu},
\end{equation}
or
\begin{equation}
f_{\mu}=   J^{\nu}B^{\nu}_{\hspace{0.08in}\mu} -
(\partial_{\mu}\widetilde{A}^{\nu})J^{\nu}.
\end{equation}
This expression matches, qua terms involved, the standard
relativistic Lorentz Force Law. It doesn't have the problem of the
extra terms that are usually present in biquaternion versions of the
Lorentz Force Law.

If charge is conserved we also have
\begin{equation}
f_{\mu}=   \partial^{\nu}T^{\nu}_{\hspace{0.08in}\mu} -
(\partial_{\mu}\widetilde{A}^{\nu})J^{\nu}
\end{equation}
as an equivalent equation. If we examen this last part
$(\partial_{\mu}\widetilde{A}^{\nu})J^{\nu}$ in more detail, an
interesting relation arises. We begin with the equation
\begin{equation}
 - \partial_{\mu} \mathcal{L} =\partial_{\mu}\widetilde{J}^{\nu}A^{\nu}.
\end{equation}
Now clearly we have
$\widetilde{J}^{\nu}A^{\nu}=\widetilde{A}^{\nu}J^{\nu}=u_0\1$ as a
Lorentz invariant scalar. Together with the chain rule this leads to
\begin{equation}
 \partial_{\mu}\widetilde{J}^{\nu}A^{\nu}=(\partial_{\mu}\widetilde{J}^{\nu})A^{\nu}+ (\partial_{\mu}\widetilde{A}^{\nu})J^{\nu} .
\end{equation}
This equation is crucial for what is to come next, the connection of
a Lagrange Equation to the Lorentz Force Law. So we have to prove it
in detail, provide an exact proof, specially because biquaternion
multiplication in general is non-commutative. We start the proof
with $\partial_{\mu}\widetilde{A}^{\nu}$:
\begin{eqnarray}
\nonumber \partial_{\mu}\widetilde{A}^{\nu}=
\left[%
\begin{array}{c}
  - \C \frac{1}{c}\partial_t \1 \\
  \nabla_1 \I\\
  \nabla_2 \J\\
  \nabla_3 \K\\
\end{array}%
\right] \left[-\C A_0 \1,
A_1 \I, A_2 \J, A_3 \K \right]=\\
\left[%
  \begin{array}{cccc}
  - \frac{1}{c}\partial_t A_0 \1         & -\C \frac{1}{c} \partial_t A_1\I   & -\C \frac{1}{c} \partial_t A_2\J & -\C \frac{1}{c}\partial_t A_3 \K  \\
  -\C \nabla_1A_0\I   & -\nabla_1 A_1\1             &  \nabla_1A_2\K           & -\nabla_1A_3\J   \\
  -\C \nabla_2A_0\J   & -\nabla_2 A_1\K             & -\nabla_2A_2\1           &  \nabla_2A_3\I   \\
  -\C \nabla_3A_0\K   &  \nabla_3 A_1\J             & -\nabla_3A_2\I           & -\nabla_3A_3\1   \\
  \end{array}
  \right]
\end{eqnarray}
In the next step we calculate
$(\partial_{\mu}\widetilde{A}^{\nu})J^{\nu}$ and use the fact that
$(\nabla_3 A_1)J_2=J_2(\nabla_3 A_1)$:
\begin{eqnarray}
\nonumber (\partial_{\mu}\widetilde{A}^{\nu})J^{\nu}= \\
\nonumber \left[%
  \begin{array}{cccc}
  - \frac{1}{c}\partial_t A_0 \1         & -\C \frac{1}{c} \partial_t A_1\I   & -\C \frac{1}{c} \partial_t A_2\J & -\C \frac{1}{c}\partial_t A_3 \K  \\
  -\C \nabla_1A_0\I   & -\nabla_1 A_1\1             &  \nabla_1A_2\K           & -\nabla_1A_3\J   \\
  -\C \nabla_2A_0\J   & -\nabla_2 A_1\K             & -\nabla_2A_2\1           &  \nabla_2A_3\I   \\
  -\C \nabla_3A_0\K   &  \nabla_3 A_1\J             & -\nabla_3A_2\I           & -\nabla_3A_3\1   \\
  \end{array}
  \right]\\
 \nonumber \left[\C J_0 \1, J_1 \I, J_2 \J, J_3 \K \right]=\\
 \left[%
  \begin{array}{c}
   (-\C\frac{1}{c}J_0\partial_t  A_0   + \C \frac{1}{c}J_1\partial_t  A_1      + \C \frac{1}{c}J_2\partial_t   A_2       + \C \frac{1}{c}J_3\partial_t  A_3)\1  \\
  (J_0\nabla_1  A_0       -J_1 \nabla_1 A_1          -J_2\nabla_1  A_2           - J_3\nabla_1  A_3)\I   \\
  (J_0\nabla_2  A_0      - J_1 \nabla_2 A_1          -J_2\nabla_2  A_2           -J_3\nabla_2  A_3)\J   \\
  (J_0\nabla_3  A_0       -J_1\nabla_3  A_1          - J_2\nabla_3  A_2          -J_3\nabla_3  A_3)\K   \\
  \end{array}
  \right]
\end{eqnarray}

Now clearly $(\partial_{\mu}\widetilde{A}^{\nu})J^{\nu}$ and
$(\partial_{\mu}\widetilde{J}^{\nu})A^{\nu}$ behave identical, only
$J$ and $A$ have changed places, so
\begin{eqnarray}
\nonumber (\partial_{\mu}\widetilde{J}^{\nu})A^{\nu}= \\
 \left[%
  \begin{array}{c}
   (-\C\frac{1}{c}A_0\partial_t  J_0   + \C \frac{1}{c}A_1\partial_t  J_1      + \C \frac{1}{c}A_2\partial_t J_2       + \C \frac{1}{c}A_3\partial_t  J_3)\1  \\
  (A_0\nabla_1  J_0       -A_1 \nabla_1 J_1          -A_2\nabla_1  J_2           - A_3\nabla_1  J_3)\I   \\
  (A_0\nabla_2  J_0      - A_1 \nabla_2 J_1          -A_2\nabla_2  J_2           -A_3\nabla_2  J_3)\J   \\
  (A_0\nabla_3  J_0       -A_1\nabla_3  J_1          - A_2\nabla_3  J_2          -A_3\nabla_3  J_3)\K   \\
  \end{array}
  \right]
\end{eqnarray}

If we add them and use the inverse of the chain rule we get
\begin{eqnarray}
\nonumber (\partial_{\mu}\widetilde{A}^{\nu})J^{\nu} + (\partial_{\mu}\widetilde{J}^{\nu})A^{\nu}= \\
 \nonumber \left[%
  \begin{array}{c}
   -\C\frac{1}{c}(\partial_t  J_0 A_0  -\partial_t  J_1 A_1  -\partial_t J_2 A_2  -\partial_t  J_3A_3)\1  \\
  (\nabla_1  J_0 A_0      - \nabla_1 J_1 A_1         -\nabla_1  J_2 A_2          - \nabla_1  J_3A_3)\I   \\
  (\nabla_2  J_0 A_0     - \nabla_2 J_1 A_1          -\nabla_2  J_2A_2           -\nabla_2  J_3A_3)\J   \\
  (\nabla_3  J_0 A_0      -\nabla_3  J_1 A_1         - \nabla_3  J_2A_2          -\nabla_3  J_3A_3)\K   \\
  \end{array}
  \right]=\\
\left[%
  \begin{array}{c}
   -\C\frac{1}{c}\1  \\
  \nabla_1 \I   \\
  \nabla_2 \J   \\
  \nabla_3 \K   \\
  \end{array}
  \right]( J_0 A_0  - J_1 A_1  - J_2 A_2  -
  J_3A_3)=\partial_{\mu}(\widetilde{J}^{\nu}A^{\nu})
\end{eqnarray}

Thus we have given the exact proof of the statement
\begin{equation}
 \partial_{\mu}\widetilde{J}^{\nu}A^{\nu}=(\partial_{\mu}\widetilde{J}^{\nu})A^{\nu}+ (\partial_{\mu}\widetilde{A}^{\nu})J^{\nu} .
\end{equation}
So we get
\begin{equation}
 -\partial_{\mu}\mathcal{L}= \partial_{\mu}\widetilde{J}^{\nu}A^{\nu}=(\partial_{\mu}\widetilde{J}^{\nu})A^{\nu}+ (\partial_{\mu}\widetilde{A}^{\nu})J^{\nu} .
\end{equation}

We now have two force equations,
$f^L_{\mu}=\partial_{\mu}\mathcal{L}=-\partial_{\mu}u_0$ and $f^T_{\mu}=-
\partial^{\nu}T^{\nu}_{\hspace{0.08in}\mu}= \frac{d}{dt}G_{\mu}$. We
combine them into a force equation that represents the difference
between these two forces:
\begin{equation}
 f_{\mu}=- f^T_{\mu} + f^L_{\mu} = \partial^{\nu}T^{\nu}_{\hspace{0.08in}\mu}+
 \partial_{\mu}\mathcal{L}.
\end{equation}
For the purely electromagnetic case this can be written as
\begin{equation}
 f_{\mu}= \partial^{\nu}\widetilde{J}^{\nu}A_{\mu}-
 \partial_{\mu}\widetilde{J}^{\nu}A^{\nu}
\end{equation}
and leads to
\begin{equation}
 f_{\mu}= (\widetilde{J}^{\nu}\partial^{\nu})
    A_{\mu}+ (\partial^{\nu}\widetilde{J}^{\nu})A_{\mu}-
 (\partial_{\mu}\widetilde{J}^{\nu})A^{\nu}- (\partial_{\mu}\widetilde{A}^{\nu})J^{\nu}.
\end{equation}
If we have $\partial^{\nu}\widetilde{J}^{\nu}=0\1$ and
$\partial_{\mu}\widetilde{J}^{\nu}=\0$ then this general force
equation reduces to the Lorentz Force Law
\begin{equation}
 f_{\mu}= (\widetilde{J}^{\nu}\partial^{\nu})
    A_{\mu}- (\partial_{\mu}\widetilde{A}^{\nu})J^{\nu}.
\end{equation}
This of course also happens if
$\partial^{\nu}\widetilde{J}^{\nu}=\partial_{\mu}\widetilde{J}^{\nu}$,
so if the RHS of this equation has zero non-diagonal terms.

\vskip12pt {\noindent
\sectionfont The Lagrangian Equation}
\vskip6pt

If the difference between $f^T_{\mu}$ and $f^L_{\mu}$ is zero, we
get the interesting equation
\begin{equation}
 - \partial^{\nu}T^{\nu}_{\hspace{0.08in}\mu}=
 \partial_{\mu}\mathcal{L}.
\end{equation}

For the situation where $\partial^{\mu}\widetilde{V}^{\nu}=0$ we
already proven the statement
\begin{equation}
 \partial^{\nu}T^{\nu}_{\hspace{0.08in}\mu}= -
 \frac{d}{dt}G_{\mu},
\end{equation}
so we get
\begin{equation}
 \frac{d}{dt}G_{\mu}=
 \partial_{\mu}\mathcal{L},
\end{equation}
which equals
\begin{equation}
 \frac{d}{dt}G_{\mu}=
 \frac{\partial \mathcal{L}}{\partial R_{\mu}}.
\end{equation}
We will prove that
\begin{equation}
 G_{\mu}= -\frac{\partial \widetilde{V}^{\nu} G^{\nu}}{\partial V_{\mu}}=\frac{\partial \mathcal{L}}{\partial
 V_{\mu}},
\end{equation}
see the Appendix for the proof and its limitations.

Combined with the forgoing equation, this leads us to
\begin{equation}
 \frac{d}{dt}(\frac{\partial \mathcal{L}}{\partial V_{\mu}})=
 \frac{\partial \mathcal{L}}{\partial R_{\mu}}
\end{equation}
as equivalent to
\begin{equation}
 - \partial^{\nu}T^{\nu}_{\hspace{0.08in}\mu}=
 \partial_{\mu}\mathcal{L}.
\end{equation}

\vskip12pt {\noindent
\sectionfont A canonical Lagrangian density}
\vskip6pt

If we choose a canonical Lagrangian density as
\begin{equation}
 \mathcal{L} = - \widetilde{V}^{\nu} G^{\nu} + \widetilde{J}^{\nu} A^{\nu}= {\bf v}\cdot{\bf g}-{\bf J}\cdot{\bf A} - u_i + \rho_e\phi  ,
\end{equation}
and an accompanying stress energy density tensor
\begin{equation}
 T^{\nu}_{\hspace{0.08in}\mu} = \widetilde{V}^{\nu} G_{\mu} - \widetilde{J}^{\nu} A_{\mu},
\end{equation}
then our force equation $f^T_{\mu}=f^L_{\mu}$ can be split in an
inertial LHS and an EM RHS
\begin{equation}
(-f^T_{\mu}+ f^L_{\mu})_{inertial}=-(-f^T_{\mu}+ f^L_{\mu})_{EM}.
\end{equation}
For situations were $(f^L_{\mu})_{inertial}=-\partial_{\mu}u_0=0$
this results in
\begin{equation}
 f_{\mu}^{inertial}= f_{\mu}^{Lorentz}.
\end{equation}
as
\begin{equation}
 \frac{d}{dt}G_{\mu}=
 J^{\nu}(\widetilde{\partial}^{\nu}
    A_{\mu})- (\partial_{\mu}\widetilde{A}^{\nu})J^{\nu}.
\end{equation}

\vskip12pt {\noindent
\sectionfont Maxwell's inhomogeneous equations}
\vskip6pt

We end with the formulation of the two inhomogeneous equations of
the set of four Maxwell Equations, as they can be expressed in our
terminology. They read
\begin{equation}
 \partial^{\nu}\widetilde{\partial}^{\nu}
    A_{\mu}- \partial_{\mu}\widetilde{\partial}^{\nu}A^{\nu}= \mu_0 J_{\mu}.
\end{equation}
As with the Lorenz Force Law, this expression matches the standard
relativistic inhomogeneous Maxwell Equations, it doesn't contain
extra terms as can be the case with the usual biquaternion
formulation of Maxwell's Equations.

The previous equation can be written as
\begin{equation}
 (-\nabla^2 + \frac{1}{c^2}\frac{d^2}{dt^2})
    A_{\mu}- \partial_{\mu}(-\partial_t \phi - \nabla \cdot {\bf A})= \mu_0
    J_{\mu},
\end{equation}
so as the difference between a wave like part and the divergence of
the Lorenz gauge part.

\vskip12pt {\noindent \sectionfont Conclusions} \vskip6pt

We have presented a specific kind of biquaternion relativistic
tensor dynamics. We formulated the general force equation
\begin{equation}\label{GFE}
 \partial^{\nu}T^{\nu}_{\hspace{0.08in}\mu}+
 \partial_{\mu}\mathcal{L}=0.
\end{equation}
The stress energy density tensor of a massive moving charged
particle in a potential field was formulated as
$T^{\nu}_{\hspace{0.08in}\mu}=\widetilde{V}^{\nu}G_{\nu}+\widetilde{J}^{\nu}A_{\nu}$
with an accompanying Lagrangian density $\mathcal{L}$ as its trace
$\mathcal{L}=T^{\nu\nu}$. Under curtain continuity conditions for the four
current and the four velocity, this leads to the Lorentz Force Law
and to the usual equations of relativistic tensor dynamics. One
advantage of our specific kind of biquaternion formalism is that it
is very akin to the standard relativistic space-time language and
that it lacks the extra terms that usually arise in biquaternionic
electrodynamics. Our formalism contains the results of both
symmetric and anti-symmetric relativistic tensor dynamics.
Curiously, our Lorentz Force Law in terms of the potentials and
currents is not anti-symmetric, nor is it symmetric. This
non-symmetric property of Eq.(\ref{LFL}) was then related to the
general force equation Eq.(\ref{GFE}).

\appendix

\vskip12pt {\noindent
\sectionfont Appendix}
\vskip6pt

We want to proof that, under curtain conditions, we have
\begin{equation}
 \frac{\partial \mathcal{L}}{\partial
V_{\mu}}= - \frac{\partial}{\partial
V_{\mu}}\widetilde{V}^{\nu}G^{\nu}=
 G_ {\nu}.
\end{equation}
The chain rule as we have used and shown before gives a first hunch.
The chain rule leads us to
\begin{equation}
 \frac{\partial}{\partial V_{\mu}}\widetilde{V}^{\nu}G^{\nu}=
 (\frac{\partial}{\partial V_{\mu}}\widetilde{V}^{\nu})G^{\nu}+ (\frac{\partial}{\partial V_{\mu}}\widetilde{G}^{\nu})V^{\nu} .
\end{equation}
As before, we cannot assume this, because it uses commutativity, so
we have to prove it.

We start the proof with $\frac{\partial}{\partial
V_{\mu}}\widetilde{V}^{\nu}$:
\begin{eqnarray}
\nonumber \frac{\partial}{\partial V_{\mu}}\widetilde{V}^{\nu}=
\left[%
\begin{array}{c}
  - \C \frac{\partial}{\partial v_0} \1 \\
  \frac{\partial}{\partial v_1} \I\\
  \frac{\partial}{\partial v_2} \J\\
  \frac{\partial}{\partial v_3} \K\\
\end{array}%
\right] \left[-\C v_0 \1,
v_1 \I, v_2 \J, v_3 \K \right]=\\
\left[%
  \begin{array}{cccc}
  - \frac{\partial}{\partial v_0} v_0 \1    & -\C  \frac{\partial}{\partial v_0} v_1\I   & -\C \frac{\partial}{\partial v_0} v_2\J  & -\C \frac{\partial}{\partial v_0} v_3 \K  \\
  -\C \frac{\partial}{\partial v_1} v_0\I   & -\frac{\partial}{\partial v_1} v_1\1       &  \frac{\partial}{\partial v_1} v_2\K     & -\frac{\partial}{\partial v_1} v_3\J   \\
  -\C \frac{\partial}{\partial v_2} v_0\J   & -\frac{\partial}{\partial v_2} v_1\K       & -\frac{\partial}{\partial v_2} v_2\1     &  \frac{\partial}{\partial v_2} v_3\I   \\
  -\C \frac{\partial}{\partial v_3} v_0\K   &  \frac{\partial}{\partial v_3} v_1\J       & -\frac{\partial}{\partial v_3} v_2\I     & -\frac{\partial}{\partial v_3} v_3\1   \\
  \end{array}
  \right].
\end{eqnarray}
Now we use the property of the orthogonal basis, so
$\frac{\partial}{\partial v_{\mu}} v_{\nu}=\delta_{\mu\nu}$:
\begin{eqnarray}
\nonumber \frac{\partial}{\partial V_{\mu}}\widetilde{V}^{\nu}= \\
\left[%
  \begin{array}{cccc}
  - 1 \1  & 0 \I     & 0\J  & 0 \K  \\
  0\I     & -1 \1       &  0\K     & 0\J   \\
  0\J     & 0\K       & -1\1     &  0\I   \\
  0\K     &  0\J       & 0\I     & -1\1   \\
  \end{array}
  \right].
\end{eqnarray}
Then we multiply $G^{\nu}$ with the result, giving
\begin{eqnarray}
\nonumber (\frac{\partial}{\partial V_{\mu}}\widetilde{V}^{\nu})G^{\nu}= \\
\nonumber \left[%
  \begin{array}{cccc}
  - 1 \1  & 0 \I     & 0\J  & 0 \K  \\
  0\I     & -1 \1       &  0\K     & 0\J   \\
  0\J     & 0\K       & -1\1     &  0\I   \\
  0\K     &  0\J       & 0\I     & -1\1   \\
  \end{array}
  \right]
  \left[\C g_0 \1, g_1 \I, g_2 \J, g_3 \K \right]=\\
\left[%
  \begin{array}{c}
   -\C g_0 \1   \\
   - g_1   \I   \\
   - g_2   \J   \\
   - g_3   \K   \\
  \end{array}
  \right] = - G_{\mu}.
\end{eqnarray}
The result of this part is
\begin{equation}
-(\frac{\partial}{\partial V_{\mu}}\widetilde{V}^{\nu})G^{\nu}=
G_{\mu}.
\end{equation}
For the second part,
\begin{equation}
-(\frac{\partial}{\partial V_{\mu}}\widetilde{G}^{\nu})V^{\nu},
\end{equation}
we have two options. The first is the easiest, assuming particle
velocity and particle momentum to be independent properties, which
makes this part zero and gives us the end result
\begin{equation}
 \frac{\partial \mathcal{L}}{\partial V_{\mu}}=-\frac{\partial}{\partial V_{\mu}}\widetilde{V}^{\nu}G^{\nu}=
 -(\frac{\partial}{\partial V_{\mu}}\widetilde{V}^{\nu})G^{\nu}= G_{\mu}.
\end{equation}
In the case that $\mathcal{L}= \widetilde{J}^{\nu}A^{\nu}$ this assumption is
allowed.

The second option is that particle velocity and particle momentum
are mutually dependent through the relation $G^{\nu}= \rho_i
V^{\nu}$, with $\rho_i$ as the inertial mass density. In that case
we have to go back to the original equation. If we assume a velocity
independent mass density this gives
\begin{eqnarray}
\nonumber \frac{\partial \mathcal{L}}{\partial
V_{\mu}}=-\frac{\partial}{\partial
V_{\mu}}\widetilde{V}^{\nu}G^{\nu}=
 -\rho\frac{\partial}{\partial V_{\mu}}(\widetilde{V}^{\nu}V^{\nu})=\\
 -\rho\frac{\partial}{\partial V_{\mu}}(v_0^2-v_1^2-v_2^2-v_3^2)= 2G_{\mu} .
\end{eqnarray}
The last situation is assumed in relativistic gravity, where the
stress energy tensor is given by $\rho_i
\widetilde{U}_{\nu}U^{\nu}$. In that situation could be tempted to
choose the Lagrangian as $\mathcal{L}=\frac{1}{2}\rho_i
\widetilde{U}^{\nu}U^{\nu}$ in order to preserve the outcome
\begin{equation}
 \frac{\partial \mathcal{L}}{\partial
V_{\mu}}= G_ {\nu}.
\end{equation}
This is done for example by Synge in his book on relativity ([8],
page 394).

But that is outside our scope. So we have to restrict the use of
\begin{equation}
 \frac{\partial \mathcal{L}}{\partial
V_{\mu}}= - \frac{\partial}{\partial
V_{\mu}}\widetilde{V}^{\nu}G^{\nu}=
 G_ {\nu}
\end{equation}
to the situations in which $V_{\mu}$ and $G_{\mu}$ are independent
of each other.

\vskip12pt {\noindent \sectionfont Acknowledgments} \vskip6pt

I am indebted to Alexander Kholmetskii and Tolga Yarman for their
support and valuable critique on earlier attempts to formulate a
biquaternion version of relativistic electrodynamics.

\vskip6pt
{\noindent \reftitlefont References}
\vskip2pt
\newcounter{Lref}
{\reftextfont
\begin{list}{[\arabic{Lref}]\hfil}
{\usecounter{Lref}\setlength{\itemsep}{0.0in}\setlength{\labelwidth}{0.3in}}

\item
A.P. Yefremov, {\it Quaternions and biquaternions: algebra, geometry, and physical theories.} arXiv: mathph/ 0501055, 2005.

\item{\label{Pauli}} W. Pauli, {\it Theory of Relativity}, Dover, New York, 1958.

\item W. Rindler, {\it Relativity. Special, General and Cosmological.},
Oxford University Press, New York, 2001.

\item A. Haas, {\it Einf\"{u}rung in die Theoretische Physik II}, Walter
de Gruyter and Co., Berlin, 1930.

\item M. von Laue, {\it Ann. Phys.}, {\bf 35}, 1911, 524-542.

\item M. von Laue, {\it Die Relativit\"atstheorie}, 6th ed., Braunschweig,
1955.

\item L. de Broglie, 1952 {\it La th\'{e}orie des particule de spin 1/2.
(\'{E}lectrons de Dirac.)}, Gauthier-Villars, Paris, 1952.

\item J.L. Synge, {\it Relativity: The Special Theory.} North-Holland
Pub. Co, Amsterdam, 2nd ed, 1965.

\end{list}
}

\end{document}